\documentclass[a4paper,11pt]{article}

\usepackage{amsmath}
\usepackage{amsfonts}
\usepackage{amssymb}
\usepackage{graphicx}
\usepackage{color}
\usepackage[footnotesize]{caption}

\newcommand{\de}{\partial}
\def\be{\begin{equation}}
\def\ee{\end{equation}}
\def\beq{\begin{equation}}
\def\eeq{\end{equation}}

\newcommand{\del}{\delta}

\newcommand{\eps}{\epsilon}
\newcommand{\zt}{\zeta}

\newcommand{\lam}{\lambda}

\newcommand{\rmd}{\mathrm{d}}
\newcommand{\nab}{\nabla}

\newcommand{\Mpl}{M_{\textrm{Pl}}}

\renewcommand{\d}{\mathrm{d}}
\newcommand{\avg}[1]{\langle #1 \rangle}
\renewcommand{\k}{\vec{k}}
\renewcommand{\O}{\mathcal{O}}
\renewcommand{\L}{\mathcal{L}}
\renewcommand{\H}{\mathcal{H}}
\newcommand{\x}{\vec{x}}

\newcommand{\grad}{\vec\nabla}
\newcommand{\kL}{k_L}
\newcommand{\kS}{k_S}
\newcommand{\vkL}{\vec k_L}
\newcommand{\vkS}{\vec k_S}

\begin{document}

\begin{center}
\Large{\textbf{The (not so) squeezed limit of the primordial 3-point function}} \\[0.5cm]
 
\large{Paolo Creminelli$^{\rm a}$, Guido D'Amico$^{\rm b}$, Marcello Musso$^{\rm a}$, Jorge Nore\~na$^{\rm c}$}
\\[0.5cm]

\small{
\textit{$^{\rm a}$ Abdus Salam International Centre for Theoretical Physics\\ Strada Costiera 11, 34151, Trieste, Italy}}

\vspace{.2cm}

\small{
\textit{$^{\rm b}$ Center for Cosmology and Particle Physics \\
Department of Physics, New York University \\
4 Washington Place, New York, NY 10003, USA}}

\vspace{.2cm}

\small{
\textit{$^{\rm c}$ Institut de Ci\`encies del Cosmos (ICC), \\
Universitat de Barcelona, Mart\'i i Franqu\`es 1, E08028, Spain}}

\vspace{.2cm}

\end{center}

\vspace{.8cm}

\hrule \vspace{0.3cm}
\noindent \small{\textbf{Abstract}\\ 
We prove that, in a generic single-field model, the consistency relation for the 3-point function in the squeezed limit receives corrections that vanish quadratically in the ratio of the momenta, i.e.~as $(k_L/k_S)^2$.  This implies that a detection of a bispectrum signal going as $1/k_L^2$ in the squeezed limit, that is suppressed only by one power of $k_L$ compared with the local shape, would rule out all single-field models. The absence of this kind of terms in the bispectrum holds also for multifield models, but only if all the fields have a mass much smaller than $H$. The detection of any scale dependence of the bias, for scales much larger than the size of the haloes, would disprove all single-field models.
  We comment on the regime of squeezing that can be probed by realistic surveys.}
\\
\noindent
\hrule

\allowdisplaybreaks

%%%%%%%%%%%%%%%%%%%%%%%%%%%%%%%%%%%%%%%%%%%%%%%%%%%%%%%%%%%%%%%%%%%%%%%

\section{Introduction and main results}

%%%%%%%%%%%%%%%%%%%%%%%%%%%%%%%%%%%%%%%%%%%%%%%%%%%%%%%%%%%%%%%%%%%%%%%
The understanding of a physical system is often enlightened by taking parametric limits. In the case of the primordial 3-point function of scalar perturbations, it is useful to consider the so-called squeezed limit, when one of the momenta in Fourier space becomes much smaller than the other two. From the theoretical point of view, it is well known that in this limit a consistency relation between the bispectrum and the power spectrum can be written, with the only assumption that perturbations are generated by a single dynamical degree of freedom  \cite{Maldacena:2002vr,Creminelli:2004yq,Cheung:2007sv}. More recently the squeezed limit has become of experimental interest, since one of the most promising signature of primordial non Gaussianity in the large scale structure, the scale-dependent bias \cite{Dalal:2007cu, Matarrese:2008nc, Slosar:2008hx}, is only sensitive to the squeezed limit of the 3-point function. Motivated by this experimental interest, in this paper we study the corrections to the consistency relation as we slightly depart from the squeezed limit.  

The consistency relation implies, in practice, the absence of terms diverging in the squeezed limit like $k_L^{-3}$, where $k_L$ is the small momentum that we send to zero in the squeezed limit. Single-field models on the market typically give a bispectrum suppressed in the squeezed limit by two powers of $k_L$, i.e.~diverging only like $1/k_L$. A natural question is therefore whether this is a general result, or also an intermediate behaviour like $1/k_L^2$ is possible in single-field models. This question is particularly relevant for measurements of the scale-dependent bias; while a behaviour $1/k_L$ is impossible to observe, being completely degenerate with the standard bias, a $1/k_L^2$ behaviour would leave interesting and potentially distinguishable signatures.

In this paper we show that it is impossible, in any single-field model, to have a $1/k_L^2$ divergence of the bispectrum in the squeezed limit. More precisely we will modestly sharpen up the consistency relation, rewriting it as
\begin{equation}
\label{eq:main}
\avg{\zeta(\k_1) \zeta(\k_2) \zeta(\k_3)} \simeq
  -(2 \pi)^3 \delta(\k_1 + \k_2 + \k_3) P(k_1) P(k_S) \left[ \frac{\rmd \ln (k_S^3 P(k_S))}{\rmd \ln k_S}  + \mathcal{O} \left(\frac{k_1^2}{k_S^2}\right)  \right] \, , \, \quad k_1 \ll k_S   \, ,
\end{equation}
where $\vec{k}_S \equiv (\k_2 - \k_3)/2$. Once the bispectrum is written in terms of $k_S$, i.e.~symmetrically with respect to the short modes $k_2$ and $k_3$, there are no linear corrections to the consistency relation. We will show this in Section \ref{sec:singlefield}, reviewing first of all the general derivation of the consistency relation, that directly leads to Eq.~\eqref{eq:main} and then, in the remaining of the Section, checking that all corrections are at least quadratic in the long wavelength momentum $k_1$. 

Our results are valid for any single-field model, with the only requirement that the unperturbed history is an attractor. No slow-roll approximation is required and indeed in Section \ref{sec:explicit} we verify explicitly the absence of linear corrections to the consistency relation in models with features or modulations in the inflaton potential. It holds also in models where scale invariance of the spectrum is not associated to a quasi de-Sitter phase, like models with a strongly varying speed of sound \cite{ArmendarizPicon:2003ht, Khoury:2008wj} or for the so-called adiabatic ekpyrosis \cite{Khoury:2009my, Khoury:2011ii}, as we discuss in Section \ref{sec:explicit}.
For simplicity, in the following, we will always call the scalar field inflaton, although our conclusions are not restricted to an inflationary scenario.

It is rather intuitive why the consistency relation only receives quadratic corrections. In a single-field model a long wavelength mode does not change the local physics -- it is adiabatic -- and in particular cannot change the short-scale 2-point function. This is true as long as we can neglect the curvature induced by this long mode, i.e.~up to order $k_L^2$. Of course what we mean by `cannot change the short scale 2-point function' depends on the gauge we choose (see for example \cite{Tanaka:2011aj}): indeed the 3-point function of $\zeta$ is not vanishing in the squeezed limit, but it is proportional to the tilt of the spectrum, since the background mode induces a (locally unobservable) rescaling of the coordinates. In this paper we show the linear corrections are completely absent if the consistency relation is written in the form \eqref{eq:main}.

Multi-field models do not satisfy any consistency relation. Indeed it is typical of multi-field models to have a local divergence of the 3-point function in the squeezed limit, i.e.~$1/k_L^3$, with a generic, potentially large parameter $f_{\rm NL}^{\rm local}$ replacing the small tilt in eq.~\eqref{eq:main}. However, in Section \ref{sec:multifield}, we argue that also in multifield models it is quite difficult to have linear corrections to the local form, i.e.~going as $1/k_L^2$. In particular we will show that the absence of this kind of term is general if all the fields participating in the generation of primordial perturbations are nearly massless, i.e.~$m \ll H$. No general conclusion can be drawn for models with fields with masses of order Hubble during inflation, as for example in the so-called quasi-single field models \cite{Chen:2009zp}. As the consistency relation applies for $k_L/k_S \to 0$, in Section \ref{sec:halobias} we quantify the regime of squeezing that can be realistically probed by measurements of the scale dependent bias. Conclusions are drawn in Section \ref{sec:conclusions}.

\section{Single-field consistency relation with subleading corrections}
%%%%%%%%%%%%%%%%%%%%%%%%%%%%%%%%%%%%%%%%%%%%%%%%%%%%%%%%%%%%%%%%%%%%%%%
\label{sec:singlefield}

In this Section we review the standard arguments leading to the consistency relation, discussing the necessary assumptions and approximations, and show how these can be partially relaxed. In the standard derivation \cite{Maldacena:2002vr, Creminelli:2004yq, Cheung:2007sv}, the long-wavelength field is assumed to be a spatially homogeneous background, and the resulting statement is therefore valid strictly when $k_1 \to 0$ and $\vec k_2 \to -\vec k_3$. We will show that one can include mild spatial variations of the background, without substantially modifying the result. %This will allow us to investigate the behavior of the three-point function when moving away from the exact squeezed limit.
The main result of this Section, and of the paper, will be to prove that
%\begin{align}
%  \avg{\zeta(\k_1) \zeta(\k_2) \zeta(\k_3)} &=
%  -(2 \pi)^3 \delta(\k_1 + \k_2 + \k_3) P(k_1) P(k_S) \left[ \frac{\rmd \ln (k_S^3 P(k_S))}{\rmd \ln k_S}  + \mathcal{O} \left(\frac{k_1^2}{k_S^2}\right)  \right] 
%\end{align}
%with $\k_S \equiv (\k_2 - \k_3)/2$, meaning that once 
the bispectrum written in terms of $\k_S = (\k_2 - \k_3)/2$ does not receive linear corrections in $k_1/k_S$ when moving away from the exact squeezed limit.

%%%%%%%%%%%%%%%%%%%%%%%%%%%%%%%%%%%%%%%%%%%%%%%%%%%%%%%%%%%%%%%%%%%%%%%
\subsection{Review of the standard leading order computation}
%%%%%%%%%%%%%%%%%%%%%%%%%%%%%%%%%%%%%%%%%%%%%%%%%%%%%%%%%%%%%%%%%%%%%%%

We review here the proof of the leading order consistency relation, following \cite{Cheung:2007sv}.
In the squeezed limit, the mode $\zeta(\k_1)$ leaves the Hubble radius much earlier than the other two, and it acts as a constant background for the other two modes.
The bispectrum can be computed in two steps: first one computes the two-point function $\avg{\zeta_2 \zeta_3}_{\bar\zeta(x)}$ in the presence of a background $\bar\zeta(x)$, and then averages this two-point function over the realizations of the background:
\begin{equation}
\label{eq:subconsistency}
  \avg{\zeta_L(\k_1) \zeta(\k_2) \zeta(\k_3)} \simeq  \avg{\bar\zeta(\k_1) \avg{\zeta(\k_2) \zeta(\k_3)}_{\bar\zeta}} \; .  
\end{equation}

We therefore need the function $\avg{\zeta(\k_2) \zeta(\k_3)}_{\bar\zeta(x)}$, which can be conveniently computed in coordinate space before taking a Fourier transform.
When the background is assumed to be homogeneous,  $\bar\zeta(x)\equiv\bar\zeta$, it can be reabsorbed simply by rescaling $\tilde x^i= e^{\bar\zeta} x^i$.
After such rescaling $\bar\zeta$ no longer appears in the action, so that the 2-point function in the new coordinates is the same as in the absence of $\bar\zeta$.
In the limit of $\bar\zeta$ constant, this reads $\avg{\zeta(\x_2)\zeta(\x_3)}_{\bar\zeta} = \avg{\zeta(\tilde\x_2)\zeta(\tilde\x_3)}$.
%Therefore, also the 2-point function $\avg{\zeta(\tilde\x_2)\zeta(\tilde\x_3)}$ will depend on the new coordinates $\tilde\x_2$ and $\tilde\x_3$ just like in the old reference frame $\avg{\zeta(\x_2)\zeta(\x_3)}$ depends on $\x_2$ and $\x_3$ when there is no background:
When $\bar\zeta$ is slowly-varying, it can be conveniently evaluated in the middle point $\x_+\equiv (\x_2+\x_3)/2$ to get $\vec{\tilde{x}}_3-\vec{\tilde{x}}_2 \simeq \x_3-\x_2 + \bar\zeta(\x_+)(\x_3-\x_2)$.
Therefore one can write the 2-point function at linear order in $\bar\zeta$ as
\be
\avg{\zeta(\x_2)\zeta(\x_3)}\vert_{\bar\zeta(x)}\simeq \xi(\x_3-\x_2)+\bar\zeta(\x_+) [(\x_3-\x_2)\cdot \grad \xi(|\x_3-\x_2|)] \;,
\ee
where $\xi(|\vec{{x}}_3-\vec{{x}}_2|) \equiv
  (2\pi)^{-3}\int \rmd^3 k_S 
  e^{i\k_S\cdot(\vec{{x}}_3-\vec{{x}}_2)} P(k_S)$. It is important to notice that corrections to this expression do not contain terms proportional to $\vec{\nab} \bar\zeta$, since there is no vector to contract with: everything is symmetric under the exchange $\x_2 \leftrightarrow \x_3$, so we cannot define a preferred direction.
It is easy to verify that corrections of order $\de^2 \bar\zeta$ will correct the final result only at $\mathcal{O}(k_L^2/k_S^2)$.
The 3-point correlation function will then read
\begin{equation}
\begin{split}
  \avg{\bar\zeta(\x_1)\zeta(\x_2)\zeta(\x_3)} &\simeq
  \avg{\bar\zeta(\x_1)\bar\zeta(\x_+)}[(\x_3-\x_2)\cdot \grad \xi(|\x_3-\x_2|)] \\
%  &\simeq \int \frac{\rmd^3\kL}{(2\pi)^3} \int \frac{\d^3\kS}{(2\pi)^3}\,
%  e^{i \vkL \cdot(\x_1 - \x_+) + i \vkS \cdot \x_-} \,
%  P_{\kL} P_{\kS}
%  \left[i \,\vkS\cdot\x_-\right]  \notag \\
  &\simeq \int \frac{\d^3\kL}{(2\pi)^3} \int \frac{\d^3\kS}{(2\pi)^3}\,
  e^{i \vkL \cdot (\x_1-\x_+)} \, P(\kL) P(\kS)
  \left[\vkS\cdot\frac{\partial}{\partial\vkS}\right]
  e^{i \vkS \cdot \x_-} \;,
\end{split}
\end{equation}
where $\x_-\equiv\x_3-\x_2$. After integrating by parts, inserting $1 = \int \d^3k_1 \,\delta(\vec k_1 + \vkL)$ and using the relation $(\partial/\partial\vkS)\cdot[\vkS P(\kS)]=P(\kS)\rmd \ln[\kS^3P(\kS)]/\rmd\ln\kS$ one gets
\begin{equation}
  -\int \frac{\d^3k_1 \, \d^3\kL \,\d^3\kS}{(2\pi)^9}\,
  e^{-i \vec k_1 \cdot \x_1 -i\vkL \cdot \x_+ + i \vkS \cdot \x_-} 
  \bigg[(2\pi)^3\delta(\vec k_1 + \vkL) P(k_1)P(\kS)
  \frac{\rmd \ln \kS^3P(\kS)}{\rmd\ln\kS}
  \bigg].
\end{equation}
In this expression, one can set $\vkL=\k_2+\k_3$ and $\vkS=(\k_2-\k_3)/2$ and get $-i\vkL \cdot \x_+ + i \vkS \cdot \x_-=-i\k_2\cdot\x_2 -i\k_3\cdot\x_3$. 
%with $\k_2=(\vkL/2)+\vkS$ and $\k_3=(\vkL/2)-\vkS$
Changing variables in the integration and Fourier transforming, one finally obtains
\begin{equation}
%  \int \frac{\d^3k_1 \, \d^3k_2 \,\d^3k_3}{(2\pi)^9}\,
%  e^{-i \vec k_1 \cdot \x_1 -i\k_2\cdot\x_2 -i\k_3\cdot\x_3}\bigg[
  \avg{\zeta_L(\k_1) \zeta(\k_2) \zeta(\k_3)} =
  -(2\pi)^3\delta(\vec k_1 + \vec k_2 + \vec k_3)
  P(k_1)P(\kS)
  \frac{\rmd \ln\kS^3P(\kS)}{\rmd\ln\kS}%\bigg],
\end{equation}
which is the leading order result of \eqref{eq:subconsistency}.

Notice that writing $\kS$ in terms of $k_2$ and $k_3$ it is straightforward to get $P(\kS)=\frac{1}{2}[P(k_2)+P(k_3)+\O(\kL/\kS)^2]$, where the linear corrections cancel out. Therefore, since the logarithmic derivative above is just the definition of the tilt of the power spectrum, the consistency relation can also be written as
%Once this is done, since $\k_1 + \k_2 + \k_3=0$ one can write $\k_S= \k_2 +\k_1/2=-\k_3 - \k_1/2$. It follows that $k_S = k_2[1+(k_1/k_2)\hat k_1\cdot \hat k_2 + \O(k_1/k_S)^2]$, with $\hat k_1$ and $\hat k_2$ unit vectors, and similarly $k_S = k_3[1+(k_1/k_3)\hat k_1\cdot \hat k_3 + \O(k_1/k_S)^2] = k_3[1-(k_1/k_2)\hat k_1\cdot \hat k_2 + \O(k_1/k_S)^2]$. Writing $P(k_S)=\frac{1}{2}[P(k_S)+P(k_S)]$ and substituting the two above expressions for $k_S$ one finds that the linear corrections cancel out to give
\begin{equation}
  \avg{\zeta(\k_1) \zeta(\k_2) \zeta(\k_3)} =
  -(2 \pi)^3 \delta(\k_1 + \k_2 + \k_3) \, 
  (n_s-1) \, P(k_1) \, \frac{P(k_2) + P(k_3)}{2} 
%  \bigg[n_S-1 + \O\bigg(\frac{k_1}{k_2}\bigg)^{\!2}\bigg]\, ,
\end{equation}
up to corrections that are quadratic in $\kL/\kS$. In this way it is explicit that the behaviour in the squeezed limit is the same at the one of a local shape up to quadratic corrections.

%%%%%%%%%%%%%%%%%%%%%%%%%%%%%%%%%%%%%%%%%%%%%%%%%%%%%%%%%%%%%%%%%%%%%%%
\subsection{A non-standard proof with standard methods}
%%%%%%%%%%%%%%%%%%%%%%%%%%%%%%%%%%%%%%%%%%%%%%%%%%%%%%%%%%%%%%%%%%%%%%%

The fact that the effect of a background mode $\bar{\zeta}$ amounts to a rescaling of the coordinates can also be seen at the level of the action.
In the exact squeezed limit, the long-wavelength field is constant in time and space: the only relevant cubic terms are thus the ones with (at least) one leg without derivatives, which will be evaluated on the long-wavelength mode.
If we focus on these terms in Maldacena's cubic action for $\zeta$  -- see eq.~(3.7) of \cite{Maldacena:2002vr} -- after some integrations by parts one ends up with
\be
\label{eq:cubicsqueezed}
S_2 + S_3 = \Mpl^2 \int \d^4 x \,\epsilon \, a^3 \left[(1+ 3 \bar\zeta) \dot\zeta^2 - (1+ \bar\zeta) (\partial_i\zeta)^2/a^2\right] \; ,
\ee
where $\bar\zeta$ is the leg that will be evaluated on the long mode and $\epsilon \equiv - \dot{H}/H^2$. In this expression we have also written the quadratic action for $\zeta$ as it makes evident how the consistency relation works: $\bar\zeta$ can be got rid of through a spatial coordinates rescaling $x^i \to x^i (1+ \bar\zeta)$ and we are left with the unperturbed quadratic action, so that the 2-point function calculation is insensitive to the background wave in the new coordinates.  Exactly the same will happen for any single field model: the same coordinate change eliminates the cubic interactions in the squeezed limit. 

Although the rescaling argument is very intuitive, for completeness we give an alternative proof using the standard computation of the 3-point function from the cubic action.
For simplicity in the remainder of the Section we focus on the simplest slow-roll inflationary scenario. For a different derivation of a very similar result see \cite{Ganc:2010ff,RenauxPetel:2010ty}.
Using the in-in formalism, the tree level bispectrum reads
\begin{equation}
  \avg{\zeta_1\zeta_2\zeta_3} =
  i \int_{-\infty}^t\d t' \avg{\left[\zeta_1\zeta_2\zeta_3,
  \int \d^3 x\, \epsilon \, a^3\zeta\left(3\dot\zeta^2
  - (\vec\nabla\zeta)^2/a^2\right)\right]}\;,
\end{equation}
where the coordinate dependence of the fields $\zeta_i\equiv\zeta(t,\x_i)$ and $\zeta\equiv\zeta(t',\x)$ is understood. Under the assumptions above, we can retain only the contractions of $\zeta_1$ (the long-wavelength mode) with $\zeta$ and neglect those with $\dot\zeta$ and $\vec\nabla\zeta$. Moreover, in the limit in which we are working, $\avg{\zeta_1\zeta}$ is just a real constant. We therefore have
\begin{equation}
  \avg{\zeta_1\zeta_2\zeta_3} =
  i \int \d^3 x \avg{\zeta_1\zeta}\avg{\left[\zeta_2\zeta_3,
  \int_{-\infty}^t\d t'\, \epsilon \, a^3\left(3\dot\zeta^2
  - (\vec\nabla\zeta)^2/a^2\right)\right]}\;.
\end{equation}

We now notice that if $\epsilon$ were not there, the integrand would be exactly the time derivative of the Hamiltonian of a free scalar field in de Sitter.
We try therefore to express the integrand in terms of the quadratic Hamiltonian and Lagrangian densities $\H_2 = a^3\epsilon \left(\dot\zeta^2 + (\vec\nabla\zeta)^2/a^2\right)$ and $\L_2 = a^3\epsilon \left(\dot\zeta^2 - (\vec\nabla\zeta)^2/a^2\right)$, in the form $[f(t)\H_2]\dot{\phantom{|}} + g(t)\L_2$.
The reason is that, once this is done, the integral of the first term simply gives $f(t)\H_2$. Since $\H_2$ generates the time evolution of the free fields, its commutator with $\zeta_2\zeta_3$ is proportional to $(\zeta_2\zeta_3)^{\,\dot{}}$ and therefore vanishes outside the Hubble radius. We will therefore only have to compute the integral of the second term, $\int_{-\infty}^t\d t' g(t')\L_2$.

The integrand can be expressed in terms of $\H_2$, $\L_2$ and their derivatives as
\begin{equation}
  a^3 \epsilon \left[3\dot\zeta^2 - (\vec\nabla\zeta)^2/a^2\right]
  = \H_2 + 2\L_2\quad;\qquad
%\end{equation}
%but also in terms of  as
%\begin{equation}
  a^3 \epsilon \left[3\dot\zeta^2 - (\vec\nabla\zeta)^2/a^2\right]
  = - \frac{1}{H}\left[{\dot \H}_2
  + \frac{\dot\epsilon}{\epsilon}\L_2 \right].
\end{equation}
Writing ${\dot\H}_2/H = (\H_2/H)\dot{\phantom{|}} - \epsilon\H_2$ and combining with the two previous expressions one gets
\begin{equation}
  a^3 \epsilon\left[3\dot\zeta^2 - (\vec\nabla\zeta)^2/a^2\right]
  = - \frac{1}{1-\epsilon}\left[\left(\frac{\H_2}{H}\right)^{\!\cdot} 
  + \left(\frac{\dot\epsilon}{H\epsilon}+2\epsilon\right)\L_2\right].
\end{equation}
The same process can be repeated to bring $1/(1-\epsilon)$ inside the total time derivative (as done with $1/H$), and iterated up to arbitrary order,
%and after one more iteration
%\begin{equation}
%  a^3 \epsilon\left[3\dot\zeta^2 - (\vec\nabla\zeta)^2/a^2\right]
%  = - \frac{1}{1-\frac{1}{H}\left(\frac{1}{1-\epsilon}\right)^{\dot{}}}
%%  \left[1-\frac{1}{H}\left(\frac{1}{1-\epsilon}\right)^{\,\dot{}}\right]^{-1}
%  \left[\left(\frac{\mathcal{H}_2}{(1-\epsilon)H}\right)^{\!\cdot} 
%  + \left(\frac{\dot\epsilon}{H\epsilon} + 2\epsilon
%  + 2\frac{1-\epsilon}{H}\left(\frac{1}{1-\epsilon}\right)^{\dot{}}
%  \right)\frac{\mathcal{L}_2}{1-\epsilon}\right].
%\end{equation}
%The process can be carried on to arbitrary order, 
so that in general one has
\begin{equation}
  a^3 \epsilon\left[3\dot\zeta^2 - (\vec\nabla\zeta)^2/a^2\right]
  = - \left[\left(\frac{1}{(1-\epsilon)H}
  + \O(\epsilon^2)\right)\H_2\right]^{\cdot}
  - \left(\frac{\dot\epsilon}{H\epsilon} + 2\epsilon
  + \O(\epsilon^2)\right)\L_2 \;.
\end{equation}

As explained above, the commutator with $\zeta_2\zeta_3$ of the integral of the first term vanishes outside the Hubble radius, and we just have to integrate the second (which also vanishes in the scale invariant limit).
In order to do this, it is convenient to express $\L_2$ as $\L_2= (a^3\epsilon \dot\zeta\zeta)\dot{\phantom{|}} - \vec\nabla\cdot(a\epsilon\zeta\vec\nabla\zeta)$, using the equation of motion for the free field. The second term can be integrated by parts to give $\vec\nabla \avg{\zeta(\x_1)\zeta(\x)}$, which can be neglected under the above assumptions. All we have to compute is thus
\begin{equation}
  \int_{-\infty}^t\!\!\d t' \left(\frac{\dot\epsilon}{H\epsilon} + 2\epsilon
  + \O(\epsilon^2)\right)(a^3\epsilon \dot\zeta\zeta)\dot{\phantom{|}}
  = \left(\frac{\dot\epsilon}{H\epsilon} + 2\epsilon
  + \O(\epsilon^2)\right)a^3\epsilon \dot\zeta\zeta 
  - \int_{-\infty}^t\!\!\d t' a^3\epsilon
  \left(\frac{\dot\epsilon}{H\epsilon}
  + 2\epsilon + \O(\epsilon^2)\right)^{\!\cdot}
  \dot\zeta\zeta\;,
\end{equation}
where the second integral is already of order $\O(\epsilon^2)$ and can be neglected. Recalling that the conjugate momentum of $\zeta$ is $\pi=2a^3\epsilon \dot\zeta$, the three-point function thus reads
\begin{equation}
  \avg{\zeta_1\zeta_2\zeta_3} = -\frac{i}{2}
  \left(\frac{\dot\epsilon}{H\epsilon} + 2\epsilon +\O(\epsilon^2)\right)
  \int \d^3 x \avg{\zeta_1\zeta(\x)}
  \avg{\left[\zeta_2\zeta_3,\pi(\x)\zeta(\x)\right]},
\end{equation}
where the fields are now all evaluated at the same time. Using the canonical commutation relations $[\zeta_i,\zeta(\x)]=0$ and $[\zeta_i,\pi(\x)]=i\delta(\x_i-\x)$ one finally has
\begin{equation}
  \avg{\zeta_1\zeta_2\zeta_3} =
  \left(\frac{\dot\epsilon}{H\epsilon} + 2\epsilon +\O(\epsilon^2)\right)
  \frac{\avg{\zeta_1\zeta_2}+\avg{\zeta_1\zeta_3}}{2}
  \avg{\zeta_2\zeta_3}\;.
\end{equation}
The proof is then completed by showing that $\dot\epsilon/(H\epsilon)=2(\epsilon-\delta)$, with $\delta\equiv\ddot\phi/(H\dot\phi)$, so that $\dot\epsilon/(H\epsilon)+ 2\epsilon = 4\epsilon - 2\delta = -(n_s - 1)$.

%%%%%%%%%%%%%%%%%%%%%%%%%%%%%%%%%%%%%%%%%%%%%%%%%%%%%%%%%%%%%%%%%%%%%%%

\subsection{Including gradients: proof of the absence of linear corrections}

%%%%%%%%%%%%%%%%%%%%%%%%%%%%%%%%%%%%%%%%%%%%%%%%%%%%%%%%%%%%%%%%%%%%%%%

We want to prove that, in any model, there are no linear corrections to the consistency relation, eq.~\eqref{eq:main}. The derivations of the consistency relation above consider only interactions in the cubic action for $\zeta$ that do not have derivatives on the background $\bar\zeta$ and, moreover, take $\bar\zeta$ as time-independent. Given that in this approximation we get eq.~\eqref{eq:main}, to conclude we need to show that all terms that we neglected give only corrections of order $\O(\kL/\kS)^2$ or higher. In doing so we have also to consider what comes from the solution of the constraints which, introducing non-local terms, could invalidate our arguments.

% was assumed to be exactly uniform and constant outside the Hubble radius.
%When trying to relax this assumption, thereby deviating from the exact squeezed limit, one must account for the space dependence of the correlation function $\avg{\zeta(\x_1)\zeta(\x)}$, as well as for corrections introduced by gradients and time derivatives of the background. Indeed, in the previous Section all terms containing $\avg{\zeta(\x_1)\vec\nabla\zeta(\x)}$ or $\avg{\zeta(\x_1)\dot\zeta(\x)}$ were discarded.
%In the following we show that all these corrections must be at least of order $\O(\kL/\kS)^2$.
%Moreover, one also needs to check that the cubic terms arising from the solution of the constraint equations (which in general lead to non-local expressions) can never produce linear corrections. This is non trivial, since the presence of inverse Laplacians in these equations could in principle produce corrections of any order, and even terms which blow up in the squeezed limit.

To summarize, we need to prove the following statements:
\begin{enumerate}
\item Corrections will arise when we take into account the time-dependence of the background mode, which above has been taken as constant. These subleading contributions are ${\cal{O}}(\kL^2)$ as we will see in Section \ref{sec:decay}. 
\item The cubic action will contain terms with a single spatial gradient acting on the background mode, $\partial_i \bar\zeta$, which naively give ${\cal{O}}(\kL)$ corrections.  However this behaviour cancels after symmetrization: the vector $\vkL$ from the derivative will be contracted in turn with $\vec k_2$ and $\vec k_3$, which are equal and opposite at leading order in the squeezed limit. 
In other words, the short-mode 2-point function has no direction -- as it is symmetric if we exchange the 2 points -- so that the background mode gradient has nothing to contract with\footnote{This argument does not work for the 3-point function of tensor modes, as in this case the polarization tensors introduce additional structure. Indeed one can check that there are linear, ${\cal{O}}(k_L)$, corrections to the consistency relation for tensor modes discussed in \cite{Maldacena:2002vr}. }. Thus corrections are ${\cal{O}}(k_L^2)$.
\item Terms in the action with time derivatives on the long-wavelength mode, $\dot{\bar\zeta}$, give ${\cal{O}}(\kL^2)$ corrections. This is a consequence of point 1, whenever $\zeta$ tends to a constant.
\item Some cubic interactions originate from solving $\delta N$ and $N_i$ through the (linear) constraint equations. In Section \ref{sec:constraints} we prove that $\delta N = {\cal{O}}(k_L^2)$, and that $N_i$ is ${\cal{O}}(k_L)$ but, following point 2 above, it gives ${\cal{O}}(k_L^2)$ corrections after symmetrizing over the short modes. 
\end{enumerate}

\subsection{The decay of the decaying mode}
\label{sec:decay}

%%%%%%%%%%%%%%%%%%%%%%%%%%%%%%%%%%%%%%%%%%%%%%%%%%%%%%%%%%%%%%%%%%%%%%%

At lowest order in derivatives, the most generic single-field quadratic action for $\zeta$ is 
\begin{equation}
\label{eq:genquadaction}
 \Mpl^2  \int \d^4x \, \frac{a^3 \epsilon}{c_s^2} \bigg[\dot\zeta^2
  - \frac{c_s^2}{a^2}(\partial_i\zeta)^2\bigg] \,,
\end{equation}
where we have allowed an arbitrarily time-dependent speed of sound $c_s$.
After defining a modified conformal time $\rmd y = (c_s/a)\rmd t$, this action reads \cite{Khoury:2008wj}
\begin{equation}
\label{eq:q2}
 \Mpl^2 \int \d y \d^3x \,q^2 \bigg[ \zeta'^2
  - (\partial_i\zeta)^2\bigg] \,,
\end{equation}
where primes stand for derivatives with respect to $y$ and $q^2\equiv a^2 \epsilon / c_s$. The equation of motion from this action is
\begin{equation}
\label{eq:geneqzeta}
  \zeta'' + 2\frac{q'}{q}\zeta' - \nabla^2\zeta = 0 \,.
\end{equation}
In the long wavelength limit, i.e.~up to terms quadratic in the gradients, this equation admits two solutions, $\zeta_k \sim$ constant and $\zeta_k\sim \int \d y (1/q^2)$. Assuming that the constant solution is dominant -- which is equivalent to requiring that the background solution is an attractor -- it is easy to realize that (approximate) scale invariance implies  (approximately) $q \propto 1/y$ \cite{Khoury:2008wj,Baumann:2011dt}\footnote{We will come back to discuss various ways to achieve this in Section \ref{sec:wierdos}.}. This means that for the second solution $\zeta \propto y^3$: the decaying mode dies off as $y^3$ in all models which give rise to an approximately scale-invariant spectrum.  This behaviour may be changed locally if we allow for a feature in the power spectrum, or being only valid on average when a periodic modulation of the power spectrum is considered. 
 In the squeezed limit of the bispectrum, the long-wavelength mode freezes ($k y \sim 1)$ much before the others and the decaying mode dies off as $y^3$ after that. Therefore, when it is the turn of the short modes to freeze its amplitude is suppressed by $(k_L/k_S)^3$, much smaller than the corrections we are interested in.  In particular, as we can neglect the decaying mode, $\dot\zeta \sim {O(k_L^2)}$ as it comes from the gradient corrections to the dominant mode. Therefore also $\dot\zeta$ terms acting on the long-wavelength mode cannot give any correction linear in $k_L/k_S$.

As a possible caveat to the logic above, one could imagine that the decaying mode, which decays after freezing, is boosted by some effect inside the horizon which makes the mode depart from the Bunch-Davies vacuum. This happens for example in the presence of sharp features of the potential. However it is easy to realize that this cannot change our conclusion, at least for sufficiently squeezed triangles. As we squeeze the triangle the out-of-the-horizon suppression $(k_L/k_S)^3$ becomes smaller and smaller. For the decaying mode to still give some effect, and in particular to give linear corrections to the consistency relation, we would need to boost without bound its amplitude before freezing. This is impossible, as at a certain point the linear perturbation theory will break down.

The quadratic action \eqref{eq:genquadaction} is only valid for models without higher derivatives. However, there are interesting models of inflation -- ghost inflation and its generalizations \cite{ArkaniHamed:2003uz,Senatore:2004rj} -- for which the spatial kinetic term has four derivatives: $(\partial_i^2\zeta)^2$. Following the logic above, with an appropriate redefinition of the time variable, one can cast the quadratic action in the form 
\begin{equation}
\label{eq:GCquadratic}
 \Mpl^2 \int \d y \d^3x \,z^2 \bigg[\zeta'^2
  - y^2(\partial_i^2\zeta)^2\bigg] \,.
\end{equation}
The $y$ dependence in front of the spatial kinetic term is inspired by the case of Ghost Inflation: in a de Sitter background, the additional derivatives of the spatial kinetic term give a factor $1/a^2 = \tau^2$ compared to the time kinetic term. Written in this way, it is easy to realize that scale invariance (and the requirement that $\zeta$ goes to a constant out of the Hubble radius) is achieved with $z \propto 1/y$. Indeed in this case one has a ``fictitious" de Sitter dilation symmetry $(\vec x, y) \to \lambda (\vec x, y)$, which implies the scale invariance of the 2-point function. Given that the discussion above does not depend on the spatial kinetic term and we have the same time dependence in front of $\zeta'^2$, the conclusion about the behaviour of the decaying mode outside the horizon remains unaltered.

\subsection{Constraints on the constraints}
\label{sec:constraints}

Let us look at the solution of the constraint equations in general models.
To compute the bispectrum, we just need the solution for the lapse $\delta N$ and the shift $N^i$ at linear order \cite{Maldacena:2002vr}.

Consider the general action for inflation
\beq
S = S_{\rm EH} + S_m = \int \frac{\Mpl^2}{2} \sqrt{h} \left[ N R^{(3)} + \frac{1}{N} \left( E_{ij} E^{ij} - E^2 \right) \right] + S_m
\eeq
where $E_{ij} = \frac12(\dot{h}_{ij} - \nab_i N_j - \nab_j N_i)$ and $S_m$ is the inflaton action.
%which also includes corrections to the Einstein-Hilbert action.

We want to understand the general structure of the linearized constraint equations:
\begin{gather}
0 = \frac{\del S}{\del N}  = \frac{\sqrt{h}}{2} \Mpl^2 \left[ R^{(3)} - N^{-2} (E^i_j E^j_i - E^2)\right] + \frac{\del S_m}{\del N} \\
0 = \frac{\del S}{\del N^i} 
= \sqrt{h} \Mpl^2 \nab_j \left[ N^{-1} (E^j_i - \del^j_i E) \right] + \frac{\del S_m}{\del N^i} \; .
\end{gather}

For single-field models, the most general expression for the inflationary Lagrangian is given by the effective theory of inflation~\cite{Cheung:2007st}.
In the comoving (or unitary) gauge, the second order Lagrangian reads:
\beq
\label{eq:EFTI}
\begin{split}
S_m =& \int \rmd^4 x \sqrt{-g} \bigg[ \Mpl^2 \dot{H} g^{00} - \Mpl^2 (3 H^2 + \dot{H}) + \frac{1}{2} M^4 (g^{00}+1)^2 \\
&+ \frac{c_1}{2} M^3 (g^{00}+1) \del E - \frac{c_2}{2} M^2 \del E^2 - \frac{c_3}{2} M^2 \del E^i_j \del E^j_i + \ldots\bigg] \; ,
\end{split}
\eeq
where the mass scale $M$ and the dimensionless coefficients $c_i$ are in general time-dependent and the dots stand for higher derivative terms we will discuss later.

In this gauge, where the scalar field is unperturbed and $h_{ij} = a^2(t) \, \exp{(2 \zeta)} \, \del_{ij}$, we have at linear order
\begin{align}
& R^{(3)} = - \frac{4}{a^2} \nab^2 \zt \; ,\\
& \del E^i_j = \dot{\zt} \del^i_j - (\de^i N_j + \de_j N^i)/2 \; ,\\
& \del E = 3 \dot{\zt} - \de_i N^i \; .
\end{align}
The constraints therefore read
\beq
\label{eq:H}
2 \del N (3 \Mpl^2 H^2 + \Mpl^2 \dot{H} - 2 M^4) + \de_i N^i \left( 2 \Mpl^2 H + c_1 M^3 \right)= -2 \Mpl^2 \frac{\nab^2 \zt}{a^2} + \dot{\zt} \left(6 \Mpl^2 H + 3 c_1 M^3 \right)  \; ,
\eeq
\be
\label{eq:momentum}
 \de_i \del N (2 H \Mpl^2 + c_1 M^3)  +  \de_i \de_j N^j (c_2 + c_3) M^2
 = \de_i \dot \zt  (2 \Mpl^2 + 3 c_2 M^2 + c_3 M^2) \; ,
\ee
where we have used that the shift must be a gradient of a scalar function, $N^i = \partial_i\psi$, as there are no vector modes.  Eq.s \eqref{eq:H} and \eqref{eq:momentum} are inhomogeneous linear equations for $\delta N$ and $\partial_i N^i$ sourced by terms proportional to $\zeta$ and $\dot\zeta$. As we discussed in the previous Section, in the long wavelength limit $\dot \zeta \sim {\cal{O}}(k_L^2)$. The equations thus imply that both $\delta N$ and $\partial_i N^i$ are ${\cal{O}}(k_L^2)$ in the long wavelength limit. This means that neither will give corrections $ {\cal{O}}(k_L)$ to the squeezed limit of the 3-point function: $\delta N$ can be neglected, while for $N^i$ --  which is ${\cal{O}}(k_L)$ -- one can apply the same argument we used for $\partial_i \zeta$ above: linear corrections cancel once symmetrized over the short modes. 

What about higher derivative terms left out from the action \eqref{eq:EFTI}? These can be built adding suitably contracted spatial derivatives $\nabla_i$ or derivatives orthogonal to the time surface $g^{0\mu} \nabla_\mu$ (in this case the upper index zero can remain unpaired.)  For example let us look at the operators
\be
 \frac{d_1}{2} (\nabla_i g^{00})^2 + \frac{d_2}{2} (g^{0\mu} \nabla_\mu g^{00})^2 + d_3 g^{0\mu} \nabla_\mu g^{00} \delta E \;,
\ee
which will generically appear when considering inflaton Lagrangians with two (or more) derivatives acting on the inflaton $\cal{L}(\phi, \partial_\mu \phi, \nabla_\mu \nabla_\nu \phi)$, as for example in \cite{Bartolo:2010bj, Creminelli:2010qf}.   
The first operator just contributes to the linearized Hamiltonian constraint with a term $-4 d_1 \nabla^2 \delta N$, which is suppressed by $k_L^2$ with respect to the others. In general additional spatial derivatives just contribute to corrections suppressed by additional powers of $k_L^2$ to the solution of the constraint equations and thus do not change our conclusions. The second operator is more interesting because it contributes to the Hamiltonian constraint with a term $- 4 (\dot d_2 + 3 H d_2) \delta \dot N - 4 d_2 \delta \ddot N$. The Hamiltonian constraint becomes a dynamical equation for $\delta N$. Of course, these higher derivative corrections must be treated perturbatively, as in any effective field theory, substituting into them the solutions of the constraints at lowest order. In this way, they cannot change the fact that in the long wavelength limit both $\delta N$ and $\partial_i N^i$ are ${\cal{O}}(k_L^2)$. The same arguments applies to the third operator proportional to $d_3$ and in general to all higher derivative operators.

% Quadratic \zeta action, with generic coefficients. P(X,\phi). \zeta ~ 1/a^3
% \dot\zeta ~ k^2
% (k_l/k_s)^3
% No slow-roll
% Modulations: Bogolubov but ok. Out of the horizon the long mode goes always as a^-3
% Models with very varying c_s . Features + non-attractor initial conditions (Lim, Easther).
% At order k^3 QM is important

\section{Explicit check in various models}
It is straightforward to check eq.~\eqref{eq:main} in various models. We did it explicitly for the standard slow-roll result of Maldacena \cite{Maldacena:2002vr} and when one considers a small deviation from unity of the speed of sound \cite{Seery:2005wm}. It is well known that, neglecting slow-roll corrections, higher-derivative terms only give equilateral shapes, i.e.~suppressed by $k_L^2$ in the squeezed limit \cite{Babich:2004gb}. The absence of linear terms should be valid also including slow-roll corrections, though the explicit check in the formulas of \cite{Chen:2006nt} is challenging. In the following two Sections we comment on two other classes of single-field models which satisfy our equation \eqref{eq:main} in an interesting way. 
\label{sec:explicit}
\subsection{Modulations and features in the inflaton potential}
Departures from slow-roll represent, together with higher derivative terms, a possible source of large non-Gaussianity. These may appear as periodic modulations of the inflaton potential \cite{Hannestad:2009yx,Flauger:2010ja} or as localized features \cite{Chen:2006xjb,Chen:2008wn,Adshead:2011bw}.

Reference \cite{Flauger:2010ja} gives an analytic expression for the bispectrum produced by models with a periodic modulation of the inflaton potential
\be
\begin{split}
 \avg{\zeta(\k_1) \zeta(\k_2) \zeta(\k_3)} 
 & \propto \frac1{k_1^2 k_2^2 k_3^2}\left[\sin\left(\frac{\sqrt{2 \epsilon_*}}{f} \log k_t/k_*\right)  \right. \\ & \left.
 +\frac{f}{\sqrt{2 \epsilon_*}}\sum_{i \neq j} \frac{k_i}{k_j} \cos\left(\frac{\sqrt{2 \epsilon_*}}{f} \log k_t/k_*\right) - \left(\frac{f}{\sqrt{2 \epsilon_*}}\right)^2 \frac{k_t (k_1^2 +k_2^2 +k_3^2)}{k_1 k_2 k_3} \sin\left(\frac{\sqrt{2 \epsilon_*}}{f} \log k_t/k_*\right) + \ldots \right] \, ,
\end{split}
\ee
where $k_t \equiv  k_1 + k_2 + k_3$ and $k_*$ is a pivot scale.
The result is obtained as an expansion in $f/{\sqrt{2 \epsilon_*}} \ll 1$, i.e.~in the limit of many oscillations per Hubble time and the dots stand for higher order terms in this expansion. Reference \cite{Flauger:2010ja} shows that this expression satisfies the consistency relation at leading order. Notice that in the squeezed limit the second term in brackets dominates, though suppressed by $f/{\sqrt{2 \epsilon_*}}$, as the first term  diverges as $1/k_L^2$.  This  behaviour of the first term, however, seems to blatantly violate our thesis. Terms subleading in $f/{\sqrt{2 \epsilon_*}}$ luckily come to our rescue: they become relevant in the squeezed limit as the argument of the trigonometric functions contains ${\sqrt{2 \epsilon_*}}/f$. Indeed one can explicitely check that the second term cancels the $1/k_L^2$ divergence at zeroth order in $f/{\sqrt{2 \epsilon_*}}$, and the cancellation also works at first order in $f/{\sqrt{2 \epsilon_*}}$, taking into account the third term in the expression. Though higher order terms have not explicitly been worked out, we expect this kind of cancellation to occur at any order.

It is important to notice that  eq.~\eqref{eq:main} applies only for ``very" squeezed triangles
\be
\frac{k_L}{k_S} \ll \frac{f}{\sqrt{2 \epsilon_*}} \ll 1 \;,
\ee
and it is not enough to require $k_L \ll k_S$.
This makes sense physically: the modes go through a resonance when they are a factor $f/{\sqrt{2 \epsilon_*}}$ shorter than the Hubble radius, i.e.~when they have a frequency comparable to the one  of the background. To deduce the consistency relation, we need the long-wavelength mode to act as a background for the others already at the time of the resonance. This is a stronger requirement than in slow-roll models, where it is enough that the long mode is frozen when the others cross the Hubble radius. When $f/{\sqrt{2 \epsilon_*}}$ is very small, the consistency relation holds only for extremely squeezed triangles. In this case it is important to check how squeezed is squeezed for a particular observable, as the relevant range of scales may not be described by our eq.~\eqref{eq:main}.
We will address this point regarding the scale-dependent bias in Section \ref{sec:halobias}.

Similar considerations apply to the case of localized features. In \cite{Adshead:2011bw} an analytic expression for the bispectrum is obtained using the generalized slow-roll approach which reads
\be
 \avg{\zeta(\k_1) \zeta(\k_2) \zeta(\k_3)}  \propto  \frac1{k_1^3 k_2^3 k_3^3} \left[I_0(k_t) k_1 k_2 k_3 + I_1(k_t) \sum_{i \neq j} k_i^2 k_j + I_2(k_t) k_t (k_1^2 +k_2^2 +k_3^2) \right] \;,
\ee
with
\be
\begin{split}
I_0 & = \int^0_{-\infty} \frac{d\tau}{\tau} G'_B(\log\tau) (k_t \tau) \sin (k_t \tau) \\
I_1 & = \int^0_{-\infty} \frac{d\tau}{\tau} G'_B(\log\tau) \cos (k_t \tau) \\
I_2 & = \int^0_{-\infty} \frac{d\tau}{\tau} G'_B(\log\tau) (k_t \tau) \frac{\sin (k_t \tau)}{k_t \tau}  \;,
\end{split}
\ee
where $G_B$ is a function of the background quantities and thus encodes the feature of the potential. Expanding $k_t$ around $2 k_S$ inside the integrals, it is straightforward to check that terms going as $1/k_L^2$ cancel among the three contributions. As in the previous case, the expansion works only when $k_L$ is sufficiently small that the mode is already frozen while the short ones are still in the vacuum and not yet perturbed by the features of the potential. 

Another situation in which the consistency relation only holds asymptotically, for very large squeezing, is with modified vacuum states; see the discussion in~\cite{Agullo:2010ws}.

\subsection{Non-standard single-field models with scale invariant spectrum}
\label{sec:wierdos}
%The limit is $k_l \to 0$ at fixed $k_s$. The decaying mode always decays in the same way.

In the presence of a time-dependent sound speed $c_s$~\cite{ArmendarizPicon:2003ht, Khoury:2008wj} or a rapidly varying equation of state~\cite{Khoury:2009my, Khoury:2011ii},
%the curvature perturbations evolve in an effective background which can be quite different from the actual scale factor behaviour.
one can obtain a scale-invariant spectrum even when the background evolution is very different from a quasi de Sitter expansion.

%This difference is not apparent at the level of the power spectrum, which is scale invariant although the slow-roll parameter $\eps$ can be large.
%In fact, for a general speed of sound, at quadratic order the action for $\zeta$ reads:
%\be
%S = \frac{\Mpl^2}{2} \int \rmd^3 x \rmd \tau z^2 \left[ (\de_\tau \zeta)^2 - c_s^2 (\de_i \zeta)^2 \right] \; ,
%\ee
%where $z \equiv a \sqrt{2 \eps}/c_s$.
%Introducing the ``sound horizon'' time $\rmd y \equiv c_s \rmd \tau$, we can write the action as
%\be
%\label{eq:nonStandardAction}
%S = \frac{\Mpl^2}{2} \int \rmd^3 x \rmd y q^2 \left[ (\de_y \zeta)^2 - (\de_i \zeta)^2 \right] \; ,
%\ee
%where $q \equiv \sqrt{c_s} z = a \sqrt{2 \eps/ c_s}$.
%In this form, the kinetic term takes the standard form as if $c_s = 1$ and the dependence on the speed of sound has been reabsorbed using a different time variable. 
In general, given an action of the form \eqref{eq:q2}, the power spectrum will be scale invariant if and only if $q''/q = 2/y^2$. To see this, write down the equation of motion for the variable $v_k \equiv q \zeta_k$ 
\begin{equation}
\partial_y^2 v_k + \bigg(k^2 - \frac{q''}{q}\bigg)v_k = 0\,,
\end{equation}
which will be the same as in de Sitter when $q''/q = 2/y^2$, with the only difference being that here the time variable is $y$.
 In general this can be realized if $q \propto 1/y$ (case I) or $q \propto y^2$ (case II). In case II the growing mode solution gives $\zeta \propto v y^{-2}$ which goes like $y^{-3}$ for modes much larger than the sound horizon $|ky| \ll 1$: the background is thus not an attractor and it is quite contrived to get a viable scenario \cite{Baumann:2011dt}. These models do not satisfy any consistency relation and we neglect them in the following.

In case I, the linear equation of motion for $\zeta$ derived from the action \eqref{eq:q2} is exactly the same as in usual slow-roll inflation with the only difference being that the time variable is $y$, and therefore one expects $\zeta$ to have a similar behaviour: it is constant outside the sound horizon $|ky| \ll 1$ with corrections being at least quadratic in $k_L$, and the decaying mode of this solution will be suppressed by $y^3$, as explained in section \ref{sec:decay}. One therefore expects all the arguments of previous sections to apply to these models, with one notable difference: scale invariance of the spectrum does not imply a scale invariant bispectrum.
These models will exhibit in general a strongly scale dependent 3-point function, which implies that they become strongly coupled after a limited ($\sim$ 10) number of efolds \cite{Baumann:2011dt}. Let us now focus on specific realizations of this case.

If one assumes that $\eps$ is constant, an expanding background has $q = - 1/y$, which for constant sound speed corresponds to de Sitter expansion $a(\tau) = - 1/\tau$.
However, a varying $c_s$ allows for sizeable departures from de Sitter expansion~\cite{ArmendarizPicon:2003ht, Khoury:2008wj} .
For instance, if $\eps = {\rm const}$, a scale invariant spectrum is obtained for $\eps_s = -2 \eps$, where $\dot{c}_s = \eps_s H c_s$. Of course, the requirement of having 
%a suitably long period of 
inflation imposes $\eps < 1$.
A stronger bound comes from the running of the gravitational wave spectrum, which does not depend on the speed of sound, and which translates into $\eps \lesssim 0.3$. Non Gaussianities can nevertheless be large and peaked in the equilateral limit since the speed of sound is small.
Since the sound speed decreases with time, and will be smaller when smaller scales cross the horizon, the bispectrum will be larger at smaller scales.

Similar considerations apply to the contracting ``adiabatic ekpyrosis'' scenario of refs.~\cite{Khoury:2009my, Khoury:2011ii}. In such a model ekpyrosis is driven by a single scalar field, and perturbations are generated during a phase in which the universe is slowly contracting, \emph{i.e.} $a(\tau) \approx \mathrm{constant}$, but the ``slow-roll'' parameter $\eps$ varies rapidly with time, $\eps \sim 1/\tau^2$. 
%This generically requires a stiff potential for the scalar field and can be realized for example in some range of field values for a potential of the form $V(\phi) = V_0(1- e^{-c\phi/M_{\mathrm{pl}}})$. 
The quadratic action for $\zeta$ is again given by an action of the form \eqref{eq:q2} with a speed of sound equal to unity and $q \propto 1/y$. Due to the fact that $\eps$ increases with time, non-Gaussianities in this model will be strongly scale-dependent and are expected to be larger at small scales. 

%It is pointed out in \cite{} that the fact that the three-point function becomes larger at the smallest scales can potentially lead to a breakdown of perturbation theory, which can be suitably fixed by changing the potential. This however generically implies that the range on which perturbations are approximately scale invariant and gaussian is only approximately a dozen e-foldings.

Even though the bispectrum is not scale-invariant, since we are dealing with single-field models the ``squeezed limit'' $k_1 \rightarrow 0$ at fixed $k_S$ is expected to be of the form
\begin{align}
  \avg{\zeta(\k_1) \zeta(\k_2) \zeta(\k_3)} &=
  -(2 \pi)^3 \delta(\k_1 + \k_2 + \k_3) P(k_L) P(k_S) \left[ \frac{\rmd \ln (k_S^3 P(k_S))}{\rmd \ln k_S}  + \mathcal{O} \left(k_L^2\right)  \right] \, .
  \label{eq:nonInvariantSqueezedLimit}
\end{align}
Due to the absence of scale invariance in the bispectrum, we cannot say that the corrections to the squeezed limit will behave as $\mathcal{O}(k_L^2/k_S^2)$, but only that they will be proportional to $k_L^2$ at fixed $k_S$. The scaling in $k_S$ will depend on the evolution of the speed of sound and ``slow-roll'' parameters with time, which fixes the scale dependence of non-Gaussianities.

One can check explicitly that, in these models, the bispectrum does indeed take the form of equation \eqref{eq:nonInvariantSqueezedLimit}.
The case of varying speed of sound is discussed in~\cite{Khoury:2008wj}
%, using the action derived in refs.~\cite{Chen:2006nt, Seery:2005wm},
and the bispectrum is given in their eq.~(7.38)\footnote{
There is a typo in the published version of the paper of their expression $\mathcal{A}^{(I)}_{\eps^2}$, which has been corrected in the latest preprint version.}.
%actually should read
%\be
%\mathcal{A}^{(I)}_{\eps^2} = - \frac{\eps^2}{64 \bar{c}_s^2} \cos\frac{\al \pi}{2} \Gamma(1+\al) (4+\al) (k_1-k_2-k_3) (k_1+k_2-k_3) (k_1-k_2+k_3) \; .
%\ee}.
Using those expressions, it can be easily verified that the corrections to the squeezed limit are quadratic in $k_L$ for each operator of the action except for $\zeta \dot\zeta^2$ and $\zeta (\partial \zeta)^2$: each one produces a bispectrum that has zeroth and linear order contributions in $k_L$, even for perfect scale invariance, but their contributions cancel.

The non-Gaussianity in the adiabatic ekpyrotic model was computed in ref.~\cite{Khoury:2011ii}. In order to compute the bispectrum, the authors perform a perturbative expansion in a parameter $1/c \ll 1$, which must be small in order to guarantee a sufficiently wide spectrum of scale-invariant perturbations. One can check in their section 5.1 that the leading contribution to the bispectrum in the squeezed limit satisfies eq.~\eqref{eq:nonInvariantSqueezedLimit}. Subleading contributions, given in their section 8.1, are not of this form in the squeezed limit. However, since they are subdominant in $1/c$ one expects the full bispectrum at subleading order to receive contributions from terms ignored in the computation of the leading piece. The total bispectrum at each order in $1/c$ is expected to satisfy our results.
%; these can come for example from subleading corrections to the mode functions and the scale factor, or even subdominant terms explicitly ignored in their final result.

\section{Multi-field case}
\label{sec:multifield}
How much of what we said above can be extended to multi-field models? Of course multi-field models do not satisfy a consistency relation for the 3-point function. Indeed, the $\delta N$ formalism \cite{Lyth:2004gb} relates, locally in space, the fluctuations of the scalar perturbations in the spatially flat gauge to the final curvature perturbation $\zeta$:
\be
\label{eq:deltaN}
\zeta(\vec x) = A_I \delta\varphi^I(\vec x) + B_{IJ} \delta\varphi^I(\vec x)\delta\varphi^J(\vec x) + \ldots
\ee
This introduces some local non-Gaussianity for $\zeta$, whose amplitude is no longer related to the tilt of the spectrum: the leading term in the squeezed limit is now a free parameter.
However, one can still wonder whether it is possible to have linear corrections.
In the following, we will show that linear corrections are absent only if one makes the additional assumption that all the fields have small mass (compared to $H$), and therefore a spectrum close to scale invariance\footnote{In eq.~\eqref{eq:deltaN}, we are implicitly assuming that the leading contribution to the 3-point function from eq.~\eqref{eq:deltaN} arises when only one of the three legs is taken at non-linear order. It is possible, however, that this contribution vanishes and one has to go to higher order in $B_{IJ}$.  For example this happens if we have two fields and $\zeta(\vec x) = A \delta\varphi_1(\vec x) + B \delta\varphi_2(\vec x)^2$ \cite{Boubekeur:2005fj};  let us check that also in this case there are no linear corrections to the $1/k_L^3$ behaviour in the squeezed limit. The bispectum can be written, assuming exact scale invariance for $\delta\varphi_2$ and with the usual definition of $k_S$, as
\be
 \avg{\zeta(\k_1) \zeta(\k_2) \zeta(\k_3)} = (2 \pi)^3 \delta(\vec k_1 + \vec k_2 + \vec k_3) \cdot 6 B^3 \Delta_{\delta\varphi_2}^3 \int \frac{d^3 q}{(2 \pi)^3} \frac1{(\vec q - \vec k_1/2)^3}\frac1{(\vec q + \vec k_1/2)^3} \frac1{(\vec q + \vec k_S)^3} \;.
\ee
In the limit $k_1 \ll k_S$ the integral is dominated by the two divergences at $\vec q = \pm \vec k_1/2$, which give a local shape with a logarithmic divergent coefficient \cite{Boubekeur:2005fj}. To study the corrections to this, it is safe to assume $q \ll k_S$, as the contribution coming from integrating outside this regime scales as $\sim 1/k_S^6$, which is irrelevant for us. Therefore, the third power spectrum can be expanded in powers of $q/k_S$ as
\be
\frac{1}{(q^2+k_S^2 - 2 \vec q \cdot \vec k_S)^{3/2}} \simeq \frac{1}{k_S^3} \left(1 - \frac32 \frac{q^2}{k_S^2} + 3 \frac{\vec q \cdot \vec k_S}{k_S^2}\right) \, .
\ee
The $\vec q \cdot \vec k_S$ term is odd if we send $\vec q \to - \vec q$, while the rest of the integrand is even, so we can drop it. The other terms just give contributions of order $k_1^2/k_S^2$ or higher.}. 

Before doing so, notice that, contrarily to the single-field case, here we have no control over the slow-roll corrections due to the deviation from de Sitter and the mass of the fields; we can only assume that they are negligible. When these corrections become large, i.e.~in the presence of fields with mass of order $H$, we cannot say anything general. Indeed, in the presence of large deviations from scale-invariance the bispectrum will contain many contributions (each with a different $k$ dependence), and it is not even well-defined what we mean by subleading corrections to the squeezed limit.

Eq.~\eqref{eq:deltaN} takes into account all the non-linearities that develop when all the modes are outside the Hubble radius. To this contribution we have to add the effect of the interaction among the fields which occur around horizon crossing, i.e.~to calculate $\langle \delta\varphi^I\delta\varphi^J\delta\varphi^K\rangle$ when the modes are comfortably outside the Hubble radius.
We want to study this correlator in the limit in which the wavevector of one of the scalars, let us call this scalar $\delta\varphi^L$,  is much longer than the others. Let us consider the action for all the $\delta\varphi$'s in spatially flat gauge and take an interaction without spatial derivatives on $\delta\varphi^L$. In the long wavelength limit we take $\delta\varphi^L$ as homogeneous and the calculation reduces, similarly to the single-field case, to the calculation of the 2-point function of the remaining modes in the presence of this background. In this case the effect of the long mode cannot be traded for a coordinate rescaling, and moreover the long mode will in general be slightly time dependent because of its small mass; for this reason also terms with a time derivative on the long mode have to be taken into account\footnote{Although the time dependence is slow-roll suppressed, we cannot neglect it as its effects pile up out of the Hubble radius and can easily become of order one.}. However what is relevant is that the effect of the long mode on the short-scale 2-point function is still proportional to the amplitude of its growing mode and therefore to the long-mode power spectrum once we average. This power spectrum will be $\sim 1/k_L^3$ and, in the squeezed limit, it adds a local contribution to the $\delta N$ calculation. 

What  we now have to check is that corrections to what we said above are all quadratic in $k_L/k_S$, similarly to what we did in Section \ref{sec:singlefield}, but now in spatially flat gauge.
\begin{enumerate}
\item Wavefunction corrections. As we are assuming that all the fields have a power spectrum close to scale-invariance, their quadratic action will be close to the one for $\zeta$, eq.~\eqref{eq:genquadaction}, but with a small mass term. Thus the only difference will be that there is a tiny time-dependence of the mode when out of the Hubble radius. Corrections to this are ${\cal{O}}(k_L/k_S)^2$ from the gradient terms and even more suppressed from the decaying mode. 
\item For terms with a spatial derivative of the long mode, the same argument as in the single-field case applies. 
\item Terms with a time derivative of the long mode have been considered above. Of course the time derivative will receive corrections from the corrections to the wavefunction, but again suppressed by at least  ${\cal{O}}(k_L/k_S)^2$ . 
\item Solutions for $\delta N$ and $N^i$. 
First of all, let us repeat the calculation of the single-field case in spatially flat gauge and then extend it to the multi-field case. The simplest way to get the constraint equations in the new gauge is to write the $\zeta$-gauge constraints -- eqs \eqref{eq:H} and \eqref{eq:momentum} -- in terms of the flat gauge variables. Using the first order relations $\zeta = - H \pi$, $N^i_{(\zeta)} = N_{\rm (flat)}^i + \partial_i \pi/a^2$ and $\delta N_{(\zeta)} = \delta N_{\rm (flat)} -\dot \pi$, we get the constraints in the new gauge
%Once we change gauge, the terms in the action \eqref{eq:EFTI}  will depend on the Goldstone of the broken time diffeomorphsims $\pi$. At second order in perturbations we have
%\begin{align}
%& g^{00} \to g^{00} (1+\dot{\pi})^2 + 2 g^{0i} \de_i \pi + g^{ij} \de_i \pi \de_j \pi \\
%& g^{0i} \to g^{0i} (1+\dot{\pi}) + g^{ij} \de_j \pi \\
%& g^{ij} \to g^{ij} \, .
%\end{align}
%We also need the expression for $\del E^i_j$ up to first order in perturbations:
%\beq
%\del E^i_j \to \del E^i_j - H \dot{\pi} \del^i_j - \de^i \de_j \pi \; .
%\eeq
%
%In order to derive the linear constraint equations, we will need the action to second order in $\delta N$ and $N^i$, but only to first order in $\pi$:
%\beq
%\begin{split}
%S_m =& \int \rmd^4 x \sqrt{h} \bigg\{ \Mpl^2 \dot{H} (- N^{-1}(1+2\dot\pi) + 2 N^i\de_i \pi) - \Mpl^2 N (3 H^2 + \dot{H}) + \frac{1}{2} M^4 (1-N^{-2}(1+2\dot\pi))^2 \\
%&+ \frac{c_1}{2} M^3 (1-N^{-2}(1+2\dot\pi)) (\del E -3 H \dot{\pi} - \nab^2 \pi/a^2)
%- \frac{c_2}{2} M^2 (\del E -3 H \dot{\pi} - \nab^2 \pi/a^2)^2 \\
%&- \frac{c_3}{2} M^2 (\del E^i_j - H \dot{\pi} \del^i_j - \de^i \de_j \pi) (\del E^j_i - H \dot{\pi} \del^j_i - \de^j \de_i \pi)  \bigg\} \;.
%\end{split}
%\eeq
%%
%We can now write down the first order hamiltonian constraint:
%\begin{multline}
%(-6 \Mpl^2 H^2 - 2 \Mpl^2\dot H  + 4M^4) \del N - (2 \Mpl^2 H  + c_1 M^3)\de_i N^i = \\
%\big(- 2\Mpl^2\dot H - 4M^4 + 3 c_1 M^3 H\big)\dot\pi + c_1 M^3 \nab^2 \pi/a^2 \; ,
%\end{multline}
\begin{multline}
2 \delta N(3 \Mpl^2 H^2 + \Mpl^2\dot H  - 2 M^4) + \de_i N^i  (2 \Mpl^2 H  + c_1 M^3)= \\
\big(2\Mpl^2\dot H - 4M^4 - 3 c_1 M^3 H\big)\dot\pi - c_1 M^3 \nab^2 \pi/a^2 - 3 (2 \Mpl^2 H \dot H + c_1 M^3 \dot H) \pi\; ,
\end{multline}
%\begin{multline}
%(2 \Mpl^2 H + c_1 M^3)\de_i \del N + (c_2 + c_3) M^2\de_i \de_j N^j = \\
%- 2 \Mpl^2 \dot{H} \de_i \pi + ( c_1 M^3 - (3c_2 + c_3)M^2 H) \de_i\dot\pi 
%- (c_2 + c_3) M^2 \de_i\nab^2 \pi/a^2\; .
%\end{multline}
\begin{multline}
(2 \Mpl^2 H + c_1 M^3)\de_i \del N + (c_2 + c_3) M^2\de_i \de_j N^j = \\
- (2 \Mpl^2 \dot{H}+  (3 c_2 + c_3) \dot H M^2 )  \de_i \pi + ( c_1 M^3 - (3c_2 + c_3)M^2 H) \de_i\dot\pi 
- (c_2 + c_3) M^2 \de_i\nab^2 \pi/a^2\; .
\end{multline}
Notice that there are two small qualitative differences compared with the $\zeta$-gauge. First of all the solution for $\delta N$ will start from a term without gradients, i.e. ${\cal O}(k_L/k_S)^0$, plus corrections which are quadratic in the gradients. This does change our conclusions, but it just says that the solutions of the constraints now contribute to interactions terms without derivatives on the long mode. The second difference is that at first sight $\partial_i N^i$ appears of order zero in the gradients. This of course cannot occur because it would imply that $N^i$ diverges as we approach an homogeneous solution. Indeed, we know from the calculation in the other gauge that this does not happen and $\partial_i N^i$ is quadratic in the gradients. 

Now let us add other fields into the game. This can be done remaining in the effective theory approach following \cite{Senatore:2010wk}. In $\zeta$-gauge, additional fields $\delta\varphi^J$ are introduced writing generic operators invariant under spatial but not time diffeomorphisms, for example $(g^{00}+1)g^{0\mu}\partial_\mu \delta\varphi^J$. Given that the $\delta\varphi$'s do not enter in the non-linear realization of the time diffeomorphisms, it is easy to see that the new operators will only contribute to the right-hand side of the constraint equations in flat gauge, with terms linear in the $\delta\varphi$'s and their derivatives. These new terms cannot change the conclusion above that $\delta N$ starts at zeroth order in the gradients with quadratic corrections, while $N^i$ must remain linear in the gradients as it must go to zero as we approach an homogeneous solution. 
\end{enumerate}
This concludes our proof that there are no linear corrections to the squeezed limit.
We have verified this in the multi-DBI models of \cite{Langlois:2008qf} and in the non-Gaussianities generated by a cubic interaction $\lam \del \varphi^3$ of an isocurvature scalar \cite{Zaldarriaga:2003my}\footnote{Notice that the formula for the bispectrum given in this work contains a typo, which has been corrected in \cite{Seery:2008qj}.}.

Notice however that we assumed from the beginning that each field has a very small mass compared to the Hubble scale during inflation. Indeed a clear counterexample to our conclusions is given by the compelling ``quasi-single field" models by Chen and Wang \cite{Chen:2009zp}. These models feature an isocurvature direction with mass {\em comparable} to $H$ and large self-coupling.
This field quickly decays out of the horizon and it features a isocurvature power spectrum which is far from scale invariance, while the power spectrum of $\zeta$ remains scale-invariant. 
However, the mixing with the adiabatic direction gives rise to a (scale-invariant) non-Gaussianity with a peculiar squeezed limit, ranging from $k_L^{-3}$, as in a local model, to $k_L^{-3/2}$, depending on the value of the mass.
In general, once large deviations from scale-invariance are allowed, one cannot derive any general property for the bispectrum.

\section{The squeezing of the scale-dependent bias.}
\label{sec:halobias}

Intuitively, since  it probes the clustering of small halos on top of large-scale density perturbations, the halo bias is sensitive to the squeezed limit of the bispectrum \cite{Dalal:2007cu, Matarrese:2008nc, Slosar:2008hx}, the regime studied in this paper.
However, as we discussed in Section \ref{sec:explicit}, for some models the consistency relation is valid only for a very small value of $k_L/k_S$.
For this reason, it is important to quantify the amount of squeezing probed by realistic measurements of the large scale bias.
In order to estimate how the halo bias depends on the bispectrum, we use the following expression derived in \cite{Matarrese:2008nc} for an analytical estimate of the correction to the halo bias due to primordial non-Gaussianity\footnote{Notice that in ref. \cite{Matarrese:2008nc} the factor $1/\mathcal{M}_R(k)$ is missing in their eq. (14).}:
\begin{multline}
\frac{\Delta b_h(k,R)}{b_h} = \frac{\delta_c}{D(z)\mathcal{M}_R(k)} \frac{1}{8\pi^2\sigma_R^2}\int_0^\infty \mathrm{d}k_1 \, k_1^2 \mathcal{M}_R(k_1)\\ \times\int_{-1}^1\mathrm{d}\mu\, \mathcal{M}_R\Big(\sqrt{k^2+k_1^2+2 k k_1 \mu}\Big) \frac{F\Big(k_1,\sqrt{k^2+k_1^2+2 k k_1 \mu},k\Big)}{P_\zeta(k)}\,,
\label{eq:bias}
\end{multline}
where $\sigma_R^2$ is the variance of the dark matter density perturbations smoothed on a scale of Lagrangian radius $R$, $\delta_c$ is the critical threshold for the collapse of a spherical object, which for a matter-dominated universe takes the value $\delta_c = 1.686$, $D(z)$ is the linear growth factor normalized to be $D(z) = (1+z)^{-1}$ during matter domination, $P_\zeta$ and $F$ are the primordial power-spectrum and primordial bispectrum for $\zeta$, and $\mathcal{M}_R$ is the linear relation between the dark matter density perturbations smoothed on a scale $R$ and the primordial curvature perturbation
\begin{equation}
\mathcal{M}_R(k) \equiv \frac{2 k^2}{5\Omega_m H_0^2}T(k) W_R(k)\,,
\end{equation}
where $\Omega_m$ is the present time fractional density of matter, $H_0$ is the present Hubble rate, $T$ is the transfer function normalized to one on large scales and $W_R$ is the filter with characteristic scale $R$, which we will take here to be top-hat in real space. 

\begin{figure}%[htc]
\begin{center}
\includegraphics[width=0.5\textwidth]{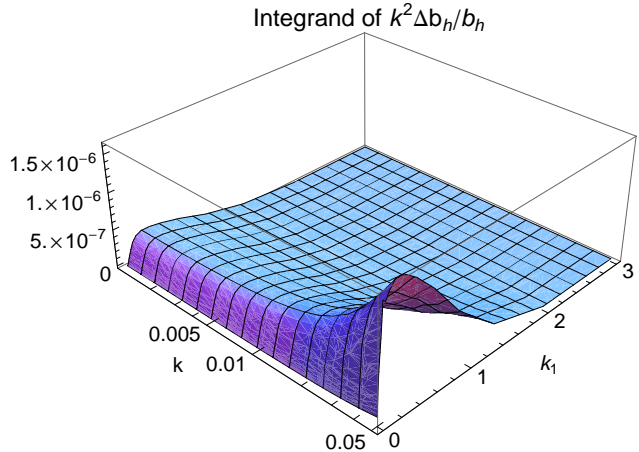}
\end{center}
\caption{\small Expression \eqref{eq:bias} for the modification to the halo bias in the presence of local non-Gaussianity multiplied by $k^2$ before integration over $k_1$, for a smoothing length of $R = 2\,h^{-1}\,\mathrm{Mpc}$ at redshift $z = 1$.}
\label{fig:integrand}
\end{figure}

Equation \eqref{eq:bias} has been shown to be in good agreement with N-body simulations \cite{Wagner:2011wx}, where several templates for the bispectrum were used to generate non-Gaussian initial conditions. Their results confirm that the scale dependent bias is sensitive to the dependence of the template on $k_L$ as it approaches the squeezed limit. Thus the equilateral template, which diverges like $1/k_L$ in the squeezed limit, has a negligible correction to the bias $\Delta b_h / b_h$ with no strong scale dependence, while the local one, which diverges like $1/k_L^3$, has a correction to the bias which grows like $1/k^2$ towards small $k$, and the orthogonal template, diverging like $1/k_L^2$, lies in between with a correction to the bias that grows roughly like $1/k$ (\footnote{We are here referring to the standard orthogonal template defined in the main text of \cite{Senatore:2009gt}. In the Appendix of the same paper another orthogonal template is defined, which goes as $1/k_L$ in the squeezed limit. See also \cite{Creminelli:2010qf}.}). This can be seen also from figure \ref{fig:integrand} where we plot $k^2 \Delta b_h(k,R)/b_h$ from equation \eqref{eq:bias} for a local bispectrum without integrating over $k_1$. Notice that the plotted function goes to a constant for small $k$ as expected, and as a function of $k_1$ it has a peak around $k_1 \sim 0.2$, a scale set by the transfer function for small objects ($R \lesssim 10\,\mathrm{Mpc}\,h^{-1}$). We thus see that the integral in equation \eqref{eq:bias} becomes large and grows as $k^{-2}$ for $k$ much smaller than the values of $k_1$ contributing the most to the integral.

\begin{figure}
\begin{center}
\includegraphics[width=0.5\textwidth]{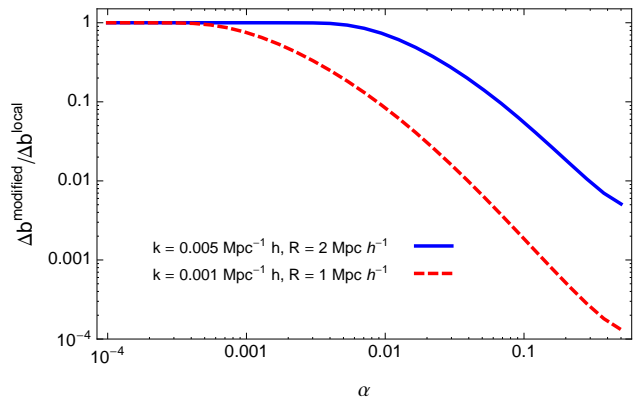}
\end{center}
\caption{\small Comparison of the non-Gaussian correction from the local template to the halo bias from a ``modified'' local template with the extreme squeezed limit removed (see text). The blue solid line is a ``conservative'' estimate for a smoothing scale corresponding to $2.5\times 10^{12} M_\odot\, h^{-1}$, the red dashed line is an ``optimistic'' estimate for larger scales and smaller objects with a mass of $3.1\times 10^{11} M_\odot\,h^{-1}$. The choice of scale for the ``optimistic'' case corresponds roughly to the largest scale accessible to the planned EUCLID survey.}
\label{fig:alpha}
\end{figure}

Using expression \eqref{eq:bias}, we can also estimate how sensitive the scale-dependent bias is to the squeezed limit, \emph{i.e.} how small a ratio $k_L/k_S$ it can probe. We do this by computing $\Delta b_h(k, R)/b_h$ for a ``modified" template that is exactly like the local one when far from the squeezed limit, but that changes its behaviour as a function of $k_L$ from $1/k_L^3$ to $1/k_L$ when $k_L \le \alpha k_S$ for a given value of $\alpha$. In other words, it is a local shape that changes to an ``equilateral'' shape when the ratio of the long mode to the short modes is less than $\alpha$. At some fixed scale $k$ and smoothing scale $R$, the correction to the bias will be insensitive to this change in the local template for very small values of $\alpha$. However, as $\alpha$ increases, we start deviating from the local case, as shown in figure \ref{fig:alpha}. In order to estimate the best we can do to probe the squeezed limit, we use a scale of $k = 0.001\, h\, \mathrm{Mpc}^{-1}$, which very roughly corresponds to the largest scales available in the planned EUCLID survey, and a smoothing length of $R = 1\, \mathrm{Mpc}\, h^{-1}$, corresponding to objects with a mass of $3.1\times 10^{11} M_\odot h^{-1}$.
From figure \ref{fig:alpha} we see that there are $\mathcal{O}(0.1)$ changes at $\alpha \approx 0.001$, so we expect the scale dependent bias to be sensitive at most to a ratio $k_L/k_S$ as small as $0.001$.
 
\section{Conclusions}
\label{sec:conclusions}
In the last few years, a large number of models giving rise to sizeable non-Gaussianities have been proposed.
In this context, it is useful to study general features which are present under very general assumptions.
Here we have made more precise the single-field consistency relation for the 3-point function, showing that there are no linear corrections $\mathcal{O}(k_L/k_S)$ as we depart from the squeezed limit, once written in the form \eqref{eq:main}.
The absence of linear corrections to the local behaviour also holds in multifield models, but only assuming that all the fields involved have a mass much smaller than $H$.
These properties are a useful check for the calculation of the bispectrum in various models, and have obvious implications for the scale dependence of the bias.
Single-field models do not give any scale-dependence in the bias, not surprisingly, as a long mode does not affect the local physics until we are sensitive to the curvature it induces, which is a ${O}(k_L^2)$ correction and corresponds to the conventional bias.
Standard multifield models (containing only very light fields) can only give a bias going as $1/k_L^2$ (associated to the local shape), while for example it is not possible to have a bias going as $1/k_L$.
However, quasi-single field models can give a different scale dependence of the bias, and it would be interesting to study how Large Scale Structure measurements can constrain these models.

\section*{Acknowledgments}
It is a pleasure to thank Xingang Chen, Justin Khoury, Enrico Pajer, Federico Piazza, Leonardo Senatore, Licia Verde, Christian Wagner and Matias Zaldarriaga for useful discussions.
The work of G. D'A. is supported by a James Arthur Fellowship.
The work of J. N. is supported by FP7- IDEAS Phys.LSS 240117.
P. C. and G. D'A. acknowledge hospitality from the Institute for Advanced Study during the completion of this project.
P. C. acknowledges hospitality from the CCPP at NYU while this work was being completed.

\end{document}